\documentclass[amsmath,amssymb,10pt,aps,prb,twocolumn,longbibliography,superscriptaddress]{revtex4-2}
\usepackage{chngcntr}
\usepackage{graphicx}
\usepackage[svgnames]{xcolor}
\usepackage[colorlinks=true,linktoc=page,linkcolor=blue,citecolor=blue,urlcolor=blue]{hyperref}
\usepackage[capitalise]{cleveref}
\usepackage{physics,mathtools,mathrsfs}
\usepackage{bm}
\def\nn{\nonumber}
\def\mr#1{\mathrm{#1}}

\begin{document}
\title{Time-resolved ARPES and optical transport properties of irradiated twisted bilayer graphene in steady-state}
\author{Ashutosh Dubey}
\thanks{These authors contributed equally to this work.}
\affiliation{Department of Physics, Indian Institute of Technology Kanpur, Kanpur 208016, India}
\author{Ritajit Kundu}
\thanks{These authors contributed equally to this work.}
\affiliation{Department of Physics, Indian Institute of Technology Kanpur, Kanpur 208016, India}
\author{Arijit Kundu}
\affiliation{Department of Physics, Indian Institute of Technology Kanpur, Kanpur 208016, India}
\begin{abstract}

We theoretically investigate the trARPES spectrum and optical Hall conductivity in periodically driven twisted bilayer graphene, considering both steady-state and ``projected" occupations of the Floquet state. In periodically driven pre-thermalized systems, steady-state occupation of Floquet states is predicted to occur when coupled to a bath, while these states have projected occupation instantaneously after the driving starts. We study how these two regimes can give markedly different responses in optical transport properties. In particular, our results show that steady-state occupation leads to near-quantized optical Hall conductivity for a range of driving parameters in twisted bilayer graphene, whereas projected occupation leads to non-quantized values. We discuss the experimental feasibility of probing such non-equilibrium states in twisted bilayer graphene.

\end{abstract}
\maketitle
\def\thefootnote{*}\footnotetext{These authors contributed equally to this work}
\section{Introduction}
The discovery of correlated insulators and superconductivity in twisted bilayer graphene (TBG) has sparked considerable theoretical and experimental interest in moiré materials \cite{Cao_2018_1,Cao_2018,Bistritzer_2011}. Moiré materials lie at the intersection of topological and electronic correlation physics, promoting novel phases that do not exist within each paradigm individually \cite{andrei2020graphene,
andrei2021marvels,
adak2024tunable,
nuckolls2024microscopic}. Near \textit{magic angles} twisted bilayer graphene feature flat bands where electron correlation effects prevail, leading to various symmetry-broken phases \cite{yankowitz2019tuning,Xie_2021,Sharpe_2019,Cao_2020,
lu2019superconductors,Polshyn_2019,liu2021tuning,xie2019spectroscopic,choi2019electronic,nuckolls2020strongly,saito2021hofstadter,das2021symmetry,wu2021chern,potasz2021exact,kang2020non,hejazi2021hybrid}. Additionally, the valley and spin degeneracies can result in competing orders, with the quantum metric and Berry curvature potentially playing a crucial role in stabilizing a specific order \cite{Abouelkomsan_2023,Hu_2019,Chen_2024,Wu_2020}.

Time-periodic fields can drive materials into exotic non-equilibrium phases \cite{Esin_2020,Esin_2021,Fausti_2011,Biswas_2020,Fazzini_2021, lindner2011floquet,Liu_2012,Shi_2024,kitamura2022floquet, hubener2017creating,D_Alessio_2015}, allowing external control to tune their band-geometric, topological \cite{oka2009photovoltaic, lindner2011floquet, kundu2014effective,Rudner_2020,Castro_2022}, and transport properties \cite{Kundu_2013, Kundu_2016, viola1, viola2, transport1,transport2, transport3,seshadri}. In particular, periodic driving can effectively make the bands flat in TBG even away from the magic angles and may induce finite Chern numbers and large Berry curvatures \cite{Li_2020}, suggesting quantized anomalous Hall transport signatures. However, unlike equilibrium systems, the nontrivial band topology in a many-body system does not always result in robust quantized transport and depends delicately on the occupation of Floquet states \cite{Seetharam_2015}.

Floquet states offer a convenient basis for describing the time evolution of periodically driven systems, similar to the Hamiltonian basis for static systems. However, the thermodynamics governing level occupation in static systems does not directly translate to driven quantum systems. The driving field induces heating and breaks the `reversibility' condition, which results in the Boltzmann distribution for equilibrium systems \cite{Bilitewski_2015,Seetharam_2015}. %This simplifies to the Fermi-Dirac distribution for non-interacting electrons. In contrast, this behavior changes under periodic driving. 
Time-resolved ARPES (trARPES) \cite{wang2013observation,Boschini_2024} and Andreev spectroscopy experiments \cite{ Park_2022} reveal the occupation of the filled Floquet band, which qualitatively differs from the Fermi-Dirac distribution observed in static systems, corroborating this distinction.

Recently, several theoretical studies have investigated the occupation of Floquet states \cite{Seetharam_2015, Fazzini_2021, iadecola2015occupation, liu2015classification}, that depends on the system-bath coupling as well as drive parameters. 
In a periodically driven closed system, the occupation of Floquet states is determined by projecting the thermal density matrix onto the Floquet states. This typically results in a finite density of excitations, with no relaxation to lower energy levels through energy emission \cite{Freericks_2009, Sentef_2015}. We refer to this as the ``projected" occupation of the Floquet states. In realistic experiments, isolating the system perfectly from the environment is challenging, and observing this kind of Floquet state occupation in practice is unlikely. On the other hand, a closed-form expression for Floquet occupation can be obtained if the system is not fully closed, the system-bath coupling is negligible, and the system reaches a steady-state on timescales longer than that of the system-bath coupling \cite{staircase, kumari2023josephsoncurrent, Shi_2024}. Generally, the excitation density is lower here than in the closed system as the system can emit energy to the bath and relax to lower levels. We refer to this occupation as the ``steady-state" occupation of the Floquet states.

In this work, we theoretically investigate the trARPES spectrum of TBG, highlighting the contrasting signatures of projected and steady-state occupations when driven by a circularly polarized pump pulse. Several studies have explored band engineering in TBG and other moir\'e systems driven by periodic fields \cite{Topp_2019, Li_2020, Rodriguez_Vega_2020, Vogl_2021, Assi_2021}. Others have examined the transport properties of periodically driven TBG, either using a semiclassical framework \cite{Yang_2023} or assuming a Fermi-Dirac occupation of Floquet states \cite{Ye_2022}. However, these studies often lack a connection between band topology and robust quantized Hall transport. We compute the optical Hall conductivity in TBG and find that when steady-state occupation is realized, the optical Hall conductivity approaches near-quantization for certain values of the drive parameters. In contrast, with projected occupation, it is never quantized. Additionally, for a generic two-band model, we find that the steady-state occupation 
is \emph{zero} (\emph{one}) for the upper (lower) quasienergy level (at zero temperature) around a gap due to a lifted degeneracy of static bands under the influence of a periodic drive. On the other hand, the projected occupation is \emph{half} for both levels. Additionally, other gaps resulting from hybridization between the quasienergy bands and Floquet sidebands lead to half-occupations of the levels in both steady-state and projected occupations.

The rest of the paper is organized as follows: in \cref{Sec:Formalism}, we describe the Floquet preliminaries (\cref{ssec:Floquet preliminaries}), occupation of the Floquet states (\cref{ssec:Occupation of Floquet mode}), trARPES spectrum (\cref{ssec:trARPES spectrum}), optical Hall conductivity (\cref{ssec:Optical Hall conductivity}) and the low energy effective model of TBG with periodic driving field (\cref{ssec:Model Hamiltonian}). In \cref{Sec:results}, we discuss the numerical results of occupation (\cref{ssec:Occupation of periodically driven TBG}), trARPES spectrum (\cref{ssec:trARPES spectrum of periodically driven TBG}), and optical Hall conductivity (\cref{ssec:Optical Hall conductivity of periodically driven TBG}) of periodically driven TBG. We conclude and summarize in \cref{sec:Summary}.

\section{Formalism}
\label{Sec:Formalism}
\subsection{Floquet preliminaries}
\label{ssec:Floquet preliminaries}
For a periodically driven quantum system, the Hamiltonian follows \( \mathcal{H}(\bm{k}, t) = \mathcal{H}(\bm{k}, t + T) \), where $\bm{k}$ represents the Bloch momentum and $T$ is the drive period. In this case, according to Floquet's theorem, the Schrödinger equation yields a complete set of orthogonal solutions of the form \( |\Psi_{\alpha}(\bm{k}, t)\rangle = e^{-i \varepsilon_{\bm{k}\alpha} t} |\phi_{\alpha}(\bm{k}, t)\rangle \). Here, $\varepsilon_{\bm k\alpha}$ is the \textit{quasienergy} of the Floquet state $|\Psi_{\alpha}(\bm{k}, t)\rangle$. \(\ket{\phi_{\alpha}(\bm k,t)}=\ket{\phi_{\alpha}(\bm k,t+T)}\) is the time-periodic part of the Floquet state, dubbed as \emph{Floquet mode}. Quasienergies lie within the first \emph{Floquet Brillouin zone}, i.e., $\varepsilon_{\bm k \alpha}\in(-\Omega/2,\Omega/2]$ correspond to unique solutions to the Schr\"odinger equation, where $\Omega = 2\pi/ T$ is the frequency of the periodic drive.
 
Due to the periodicity of \(\mathcal H(\bm k,t)\) and \(|\phi_{\alpha}(\bm k,t)\rangle\), they can be expressed in a Fourier series:

\begin{align}
   &\mathcal H(\bm k,t) = \sum_{p \in \mathbb Z} e^{-i p \Omega t}\mathcal H^{(p)}(\bm k),
   \label{eq:ham-fourier}
   \\
   &|\phi_{\alpha}(\bm k,t)\rangle = \sum_{p \in \mathbb Z} e^{-i p \Omega t}|\phi_{\alpha}^{(p)}(\bm k)\rangle. 
 \label{eq:three}
\end{align}
$\mathcal H^{(p)}(\bm k)$ and 
$|\phi_{\alpha}^{(p)}(\bm k)\rangle$ denotes the $p$-th Fourier coefficient of $\mathcal{H}(\bm{k}, t)$ and $|\phi_{\alpha}(\bm{k}, t)\rangle$ as $\mathcal{H}^{(p)}(\bm{k})$ and $|\phi^{(p)}_\alpha(\bm{k})\rangle$, respectively. 

Substituting Eq.~\eqref{eq:ham-fourier} and Eq.~\eqref{eq:three} into the Schr\"{o}dinger equation, yields the following eigenvalue equation:
\begin{equation}
   \sum_{n} [\mathcal H^{(m-n)}(\bm k)- m\Omega\delta_{m,n}]|\phi_{\alpha}^{(n)}(\bm k)\rangle = \varepsilon_{\bm k\alpha}|\phi_{\alpha}^{(m)}(\bm k)\rangle.
   \label{eq:ex-ham}
\end{equation}
This defines the \emph{extended-zone} Hamiltonian $\mathbb H$, the blocks of which are given by: $\mathbb H_{m,n} = [\mathcal H^{(m-n)}(\bm k)- m\Omega\delta_{m,n}]$.
Upon diagonalizing $\mathbb H$, one finds the quasienergies ($\varepsilon_{\bm k\alpha}$)
and the Fourier components of the Floquet mode $|{\phi^{(p)}_{\alpha}(\bm k)}\rangle$. Note that the Fourier coefficient of the Floquet modes obeys the following normalization condition: $\sum_p \langle{\phi^{(p)}_{\alpha}(\bm k)}|{\phi^{(p)}_{\alpha}(\bm k)}\rangle=1$. 

\subsection{ Occupation of Floquet states}
\label{ssec:Occupation of Floquet mode}
For a driven system, the `reversibility' condition breaks down, which gives rise to the Boltzmann distribution and the  Fermi-Dirac distribution for a non-interacting static system. Consequently, the thermodynamic principles applied to a static system cannot directly translate to periodically driven systems. The occupation of Floquet states depends on the specifics of the driving protocol and system-bath coupling, which provide a relaxation mechanism to the system.

For a closed, periodically driven system, the occupation of $\alpha$th Floquet state (with Floquet mode $|\phi_{\alpha}(\bm k,t)\rangle$), for a non-interacting system is obtained by projecting the density matrix to the Floquet state,
\begin{align}
    n^{\rm pr}_{\alpha}(\bm k) = \langle \phi_{\alpha}(\bm k,0)| \rho^{\rm static}(\bm k) |\phi_{\alpha}(\bm k,0)\rangle.
    \label{eq:quasi-steady-state-occu}
\end{align}
This is referred to as the projected occupation throughout the remainder of the paper.
Here, $\rho^\text{static}(\bm k)$ is the thermal density matrix of a static system, given by
\begin{align}
    \rho^{\rm static}(\bm k) = \sum_n f(E_{\bm k n})\ketbra{\psi_n(\bm k)},
    \label{eq:thermal-dens-mat}
\end{align}
where $|\psi_n(\bm k)\rangle$ and $E_{\bm k n}$ are the eigenstate and corresponding eigenvalue for the $n$th band, respectively. $f(x) = 1/\left(e^{\beta (x-\mu) } + 1\right)$ is the Fermi-Dirac distribution function. $\beta$ and $\mu$ are the inverse temperature and the chemical potential, respectively.
The periodic drive gives rise to a finite excitation density, and the system can not relax to a lower energy level through emission as it is closed from the environment.

A closed-form expression for Floquet occupation can be obtained if the system is not fully isolated, the system-bath coupling is negligible, and the system reaches a steady-state on timescales longer than that of the system-bath coupling. In this scenario, the occupation of the $\alpha$th Floquet state follows a staircase Fermi-Dirac distribution \cite{staircase,kumari2023josephsoncurrent} given by
\begin{align}
    n^{\rm ss}_{\alpha}(\bm k) = \sum_{p\in \rm \mathbb{Z}}f(\varepsilon_{\bm k\alpha} + p\Omega - \mu)\langle \phi_{\alpha}^{(p)}(\bm k)| \phi_{\alpha}^{(p)}(\bm k)\rangle.
    \label{eq:steady-state-occu}
\end{align}
This is referred to as the steady-state occupation. In this case, the excitation density is generally lower, as the system can relax to a lower energy level.

In the following sections, we demonstrate that the realization of either projected or steady-state occupation in a periodically driven system leads to contrasting trARPES spectroscopy (\cref{ssec:trARPES spectrum of periodically driven TBG}) and optical Hall conductivity (\cref{ssec:Optical Hall conductivity of periodically driven TBG}) in twisted bilayer graphene (TBG), as well as in Dirac and semi-Dirac systems (see  \cref{app:B}). We analytically calculate both the projected and steady-state occupation probabilities for weak driving at various drive-induced gap opening points (for more details, see \cref{ssec:Occupation of periodically driven TBG,app:C}). 

\subsection{trARPES spectrum}
\label{ssec:trARPES spectrum}

 In trARPES spectroscopy, the system undergoes illumination by two ultrafast laser pulses: the \emph{pump} and \emph{probe} pulses, with a tunable delay time between them \cite{Boschini_2024}. The pump pulse drives the system out of equilibrium. The probe pulse then forces the system to eject photoelectrons, which are then captured by a photodetector with an angular resolution. From the trARPES spectrum, one can directly observe the occupied states in the quasienergy spectrum of the driven system.
 The pump pulse is modeled by a vector potential, \(\bm A_{\mathrm{pump}}(t) = A_0
s_{\mathrm{pump}}(t)(\cos(\Omega t), \sin(\Omega t))\), which is incorporated in the Hamiltonian by Peierls substitution. Here \(s_{\mathrm{pump}}(t) = e^{-(t-t_p)^2 / 2 \sigma^2_{\mathrm{pump}}} \) is a Gaussian envelop of the pump pulse with a width of \( \sigma_{\mathrm{pump}} \), centered around time \( t_p \). The spectrum of the trARPES represents the photoemission intensity $\mathcal{I}(\bm k,\omega)$ at momentum $\bm k$ and frequency $\omega$, given by \cite{Freericks_2009}
\begin{align}
	\mathcal{I}(\bm k,\omega) &=  \Im \frac{1}{(2t_{0})^{2}}\int_{-t_{0}}^{t_{0}}\int_{-t_{0}}^{t_{0}} \dd t_1 \dd t_2 \Tr[\mathcal G^{<}(\bm k, t_1, t_2)] \nn \\
	&\times s_{\mathrm{probe}}(t_1) s_{\mathrm{probe}}(t_2)e^{i \omega(t_1 - t_2)},
	\label{eq:occupation}
\end{align}
where \( s_{\mathrm{probe}}(t) = e^{-(t-t_{\mathrm{pr}})^2 / 2 \sigma^2_{\mathrm{probe}}} \) accounts for the probe pulse 
centered around time \( t_{\mathrm{pr}} \) with width of \( \sigma_{\mathrm{probe}}\). The probing interval is in the time interval $[-t_0,t_0]$. 
\( \mathcal G_{\alpha\beta}^{<}(\bm k, t,t') = i \expval{c^{\dagger}_{\bm k\beta}(t') c_{\bm k\alpha}(t)}\) is the ``lesser" Green's function \cite{kadanoff2018quantum}, where \( c_{\bm k \alpha}\) is the fermionic annihilation operator at momentum $\bm k$ for $\alpha$ orbital. The evolution of the annihilation operator \( c_{\bm k \alpha}\) is given by, 
    \begin{align}
       c_{\bm k \alpha}(t) &=  \sum_{\gamma}  [\mathcal U_{\bm k}(t,0)]_{\alpha\gamma}c_{\bm k \gamma}(0),
        \label{eq:anh}\\
        \mathcal U_{\bm k}(t,0) &
        = \mathscr T \exp(-i \int_0^t \dd t' \mathcal H(\bm k, t')),
        \label{eq:evol-op}
    \end{align}
where \(\mathcal U_{\bm k}(t,0)\) is the evolution operator in the presence of the pump pulse and $\mathscr T$ is the time ordering operator.
 The pump-probe delay time is given by \(\Delta t = t_{\mr{pr}} - t_{\mr p}\). For the peak field strength, i.e., when the pump pulse overlaps maximally with the probe pulse overlap $\Delta t = 0$, which is the case we assume throughout this work.

To observe well-defined Floquet band structures in the trARPES spectrum, it was shown in Ref. \onlinecite{Sentef_2015} that one has to follow the time scale hierarchy between the duration of pump-pulse, duration of the probe-pulse, and the time period of the drive: \( \sigma_{\mathrm{pump}} \gg \sigma_{\mathrm{probe}} \gg  T \).

\subsection{Optical Hall conductivity}
\label{ssec:Optical Hall conductivity}

When subjected to periodic driving, a system's physical properties can vary significantly from its static counterpart. Studying the properties of such systems often involves analyzing their response to weak external perturbations. Linear response theory provides a natural framework for this. Ref.~\cite{Kumar_2020} extends the linear response theory from static systems to strongly driven Floquet systems under the influence of a weak external probe, which we employ to compute Optical Hall conductivity.

trARPES spectroscopy exclusively captures filled quasienergy states, lacking information about unfilled states. However, introducing a weak perturbation typically induces optical transitions between filled and unfilled states. Consequently, optical conductivity can probe the unoccupied quasienergy states.

The formula for the optical Hall conductivity $\sigma_{xy}(\omega)$ probed at frequency $\omega$, for a generic periodically driven system with driving frequency $\Omega$ reads:
\begin{align}
 \label{eq:optical-hall}
	\mathcal \sigma_{xy}(\omega) &= \frac{1}{\omega V} \sum_{\mathclap{\substack{\alpha,\beta ,m,\bm k}}}\Im\left[\frac{D^{(m)}_{\alpha x\beta}(\bm k)D^{(-m)}_{\beta y\alpha}(\bm k)(n_{\alpha}(\bm k) - n_{\beta}(\bm k))}{\omega  + \varepsilon_{\alpha}(\bm k) - \varepsilon_{\beta}(\bm k) + m\Omega + i\delta }\right].
\end{align}
Here \(D^{(m)}_{\alpha j
\beta}(\bm k) = \frac{1}{ T}\int_{0}^{T} \dd t~ e^{i m \Omega t}\matrixel{\phi_{\bm k \alpha}(t)}{\frac{\partial \mathcal H(\bm k, t)}{\partial k_{j}}}{\phi_{\bm k \beta}(t)}\) represents the Fourier component of the matrix element of the current operator in $j$-direction. $\varepsilon_{\alpha}(\bm k)$ and $n_{\alpha}(\bm k)$ are the quasienergy and occupation probability of $\alpha$-th Floquet state, respectively, $\delta$ is an infinitesimally small positive number, and $V$ is the total area of the system. We refer the reader to \cref{eq:optical-hall} for the details of the derivation of \cref{app:A}. Note that \cref{eq:optical-hall} is valid for only insulators; for metals, there will be additional corrections corresponding to Drude conductivity \cite{floquetDrude}. 
Physically, $\sigma_{xy}(\omega)$ encapsulates all optical transitions from a filled to an unfilled quasienergy band, including sidebands separated by energy $\omega$.

In a periodically driven system, circularly polarized light explicitly breaks time-reversal symmetry and induces non-trivial Chern number.
However, in periodically driven systems, this does not always result in a quantized Hall conductivity. The occupation probability can deviate significantly from the static Fermi-Dirac distribution, causing the correspondence between the Chern number and the DC (i.e. \(\omega \to 0\)) Hall conductivity to break down at zero temperature. Nevertheless, if the occupation probability closely approximates the Fermi-Dirac distribution, near-quantized Hall conductivity can be achieved.
\begin{figure*}[ht]
\centering
\includegraphics[width=1.0\linewidth]{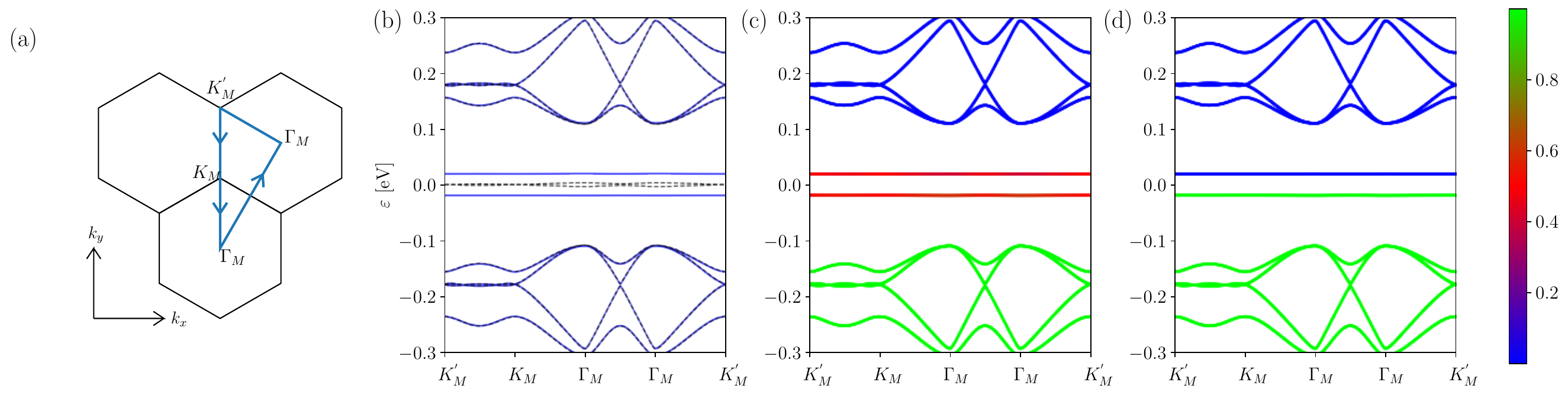}
\caption{(Color online) (a) Path along the high symmetry points in the moiré Brillouin zone of the twisted bilayer graphene (TBG). (b) Band structure of the static TBG (dashed line) and driven TBG (solid line). (c) Projected (PR) occupation and (d) steady-state (SS) occupation of TBG, with the values represented by the color bar: blue indicates zero, green indicates 1, and red indicates 1/2. The parameters used for the above plots are \(\Omega = 4.5\)~eV and \(A_0 = 1.5k_{\theta}\),  corresponding to an electric-field magnitude of $E \approx \Omega A_0 = 2.2 \times 10^4$~kV/cm. The twist angle is $\theta=1.1^\circ$ and the chemical potential is $\mu = 0$.}
\label{fig:occupation}
\end{figure*}
\subsection{Model Hamiltonian}
\label{ssec:Model Hamiltonian}
Twisted bilayer graphene (TBG) consists of two graphene layers rotated relative to each other, inducing a moiré potential due to the rotational misalignment. The low energy effective Hamiltonian is given by  \cite{Bistritzer_2011}:
\begin{align}
	\label{eq:hamiltonian}
	\mathcal H_0 = \begin{pmatrix}
		h_{\theta / 2} &  U(\bm x) \\
		 U(\bm x)^{\dagger} & h_{-\theta / 2}
	\end{pmatrix},
\end{align}
where \( h_{\pm \theta / 2}= -i  v_F \bm \nabla \cdot \bm \sigma_{\pm\theta / 2} \). 
Here, \( v_F \) is the Fermi velocity of an electron in a single graphene layer, and \( \bm \sigma_{\theta} =
e^{-i \sigma_z \theta / 2} \bm \sigma e^{i \sigma_z \theta / 2}
\) are the rotated Pauli matrices. \( U(\bm x) \) is the moir\'e potential, given by
\begin{align}
    &U(\bm x) = U_0 + U_1 e^{i \bm G_1 \cdot \bm x} + 
    U_2 e^{i \bm G_2 \cdot \bm x}, \\
    &U_n = w_{AA} \sigma_0
    + w_{AB} [\sigma_x \cos(n \phi)
    +\sigma_y \sin(n \phi)
    ], 
\end{align}
where $\phi = 2\pi / 3$ and $\bm G_{1,2} = k_{\theta}(\sqrt 3 / 2, \pm 3 / 2)$ are the reciprocal lattice vectors, and \(
k_{\theta} =  8\pi \sin(\theta / 2) / 3 a \), with $a$ as the lattice constant of graphene. Here, \( w_{AA} \) and \( w_{AB} \) are the tunneling amplitudes between AA and AB-stacked regions of TBG, respectively. The Hamiltonian in the Eq.~\eqref{eq:hamiltonian} acts on the spinor space \( (\psi_1,\chi_1,\psi_2,\chi_2) \), where \( 1,2 \) refers to two graphene layers and \( \psi_i,\, \chi_i \) corresponds two sublattices of graphene. 

We periodically drive TBG with a circularly polarized light, represented by the vector potential  \( \bm A(t) = A_0(\cos(\Omega t), \sin(\Omega t)) \), where $\Omega$ is the driving frequency. This enters in the Hamiltonian by a minimal substitution, i.e., 
\begin{align}
    h_{\pm \theta / 2} \to h_{\pm \theta / 2}(t) =   v_F [-i \bm \nabla + \bm A(t) ]\cdot \bm \sigma_{\pm\theta / 2}.
\end{align}
We treat the moir\'e potential in an effective long-wavelength approximation \cite{Li_2020}. The time-dependent Hamiltonian reads
\begin{align}
    \mathcal H(t) = \mathcal H_0 + h(t),
\end{align}
where the time-dependent part $h(t)$ is given by 
\begin{align}
    h(t) = \mathcal H^{(+1)} e^{-i \Omega t} + \mathcal H^{(-1)}e^{i \Omega t},
\end{align}
Here, $\mathcal H^{(\pm1)}$ are the Fourier components of $\mathcal H(t)$, given by
\begin{align}
	\left[\mathcal H^{(-1)}\right] &= \left[\mathcal H^{(+1)}\right]^{\dagger} \nn \\ 
	&=  v_F A_0 
	\left(
		\begin{array}{cccc}
			0 & 0 & 0 & 0 \\
			e^{ {i \theta } / {2}} & 0 & 0 & 0 \\
			0 & 0 & 0 & 0 \\
			0 & 0 & e^{{-i \theta} / {2} } & 0 \\
		\end{array}
	\right),
\end{align}
and the zeroth Fourier component is given by the static Hamiltonian, i.e., $\mathcal H^{(0)} = \mathcal H_0$. As the low energy Hamiltonian is constructed from two Dirac Hamiltonian and does not incorporate the non-linear dispersion of graphene at higher energies, only the Fourier harmonics \(\mathcal H^{(0)}, \mathcal H^{(\pm 1)} \) of $\mathcal H(t)$ are non-zero. Now, to find the quasienergy and the Fourier components of the Floquet mode, we solve the eigenvalue problem of extended-zone Hamiltonian Eq.~\eqref{eq:ex-ham}. %\cref{eq:ex-ham}.

In this study, we adopt natural units $ e = \hbar = 1 $, with additional parameters specified as follows: $ a = 2.4 $ \AA, $ v_F / a = 2.425 $ eV, $ w_{AB} = 112 $ meV, and $ w_{AA} = 0 $ meV (chiral limit).
For the construction of the Hamiltonian in plane-wave basis, the reciprocal lattice vectors \(\bm G_{n,m} = n \bm G_{1} + m \bm G_{2} \) are restricted within \( -3 \leq m,n \leq 3 \). The number of Fourier components that 
for the construction of the extended-zone Hamiltonian is restricted to seven, enough for the convergence of low-energy quasienergy bands.

\section{Results and Discussion}
\label{Sec:results}
\subsection{Occupation of periodically driven TBG}
\label{ssec:Occupation of periodically driven TBG}

When a static system is periodically driven (circularly polarized light) at a frequency much higher than the relevant bandwidth, the quasienergy spectrum resembles the real energy spectrum of the static system, except for gap openings at zero quasienergy, when there are degeneracy at zero energy in the static bands. These gap openings are attributed to broken time-reversal symmetry of the system in the presence of periodic drive. At these gap openings, interestingly, steady-state occupation predicts complete filling (assuming at zero temperature and chemical potential at zero quasienergy) of the lower quasienergy band and the upper quasienergy band to remain empty. This starkly contrasts with the projected occupation's prediction that both the upper and lower quasienergy bands will be exactly half-filled around the gap opening. A derivation of the above is presented in \cref{app:C}.

On the other hand, at lower values of driving frequency (lower than the bandwidth of static system) and for very weak driving amplitude, the quasienergy spectrum resembles a folded static energy spectrum within a Floquet Brillouin zone $(-\Omega/2,\Omega/2]$, except gap openings at zero (in addition to the gaps that forms out of broken time-reversal symmetry) as well as at quasienergy $\pm \Omega/2$. This is due to the mixing of bands as they fold. Interestingly, whenever a gap opens because of folded bands mix, both SS and PR occupation predict exactly half occupation of the quasienergy bands (see \cref{app:C}). The above analysis is true only for weak driving amplitude, and at arbitrary driving intensity, SS and PR occupations can differ significantly at all parts of the spectrum.

In the case of twisted bilayer graphene (TBG), around twist angle \(\theta = 1.1^{\circ}\), in the chiral limit, the two central bands are almost flat (bandwidth \(\sim 3.5~\mathrm{meV}\)). These bands are degenerate at the two Dirac points in the moir\'e Brilloin zone (mBZ), i.e., $K_M$ and $K'_M$ (see \cref{fig:occupation}(a)) due to the $C_2 \mathcal T$ symmetry present in graphene, where $C_2$ is $180^\circ$ rotation and $\mathcal T$ is the spinless time-reversal symmetry operator. 
Time-reversal symmetry is broken when a circularly polarized light is applied, and the quasienergy spectrum gap out at the $K_M$ and $K'_M$ points in the mBZ. This results in two gapped central quasienergy bands that are even flatter, with a bandwidth of approximately \(0.28~\mathrm{meV}\), as shown in Fig.~\ref{fig:occupation}(b). Consequently, the occupation in the quasienergy bands is redistributed, differing from that of the static bands. For these calculations, we set \(\mu = 0\) to determine the occupation of the quasienergy bands.
At zero temperature, the value of PR occupation is nearly half for the two flat bands (above and below \(\mu =0\)) as shown in Fig.~\ref{fig:occupation}(c), whereas the value of SS occupation is nearly one and zero for flat bands above and below \(\mu =0\) respectively as shown in Fig.~\ref{fig:occupation}(d). 
 Additionally, we present the PR and SS occupations for Dirac and semi-Dirac systems in \cref{app:B}.

\subsection{trARPES spectrum of periodically driven TBG}
\label{ssec:trARPES spectrum of periodically driven TBG}
  \begin{figure*}[htpb]
    \centering
    \includegraphics[width=0.7\linewidth]{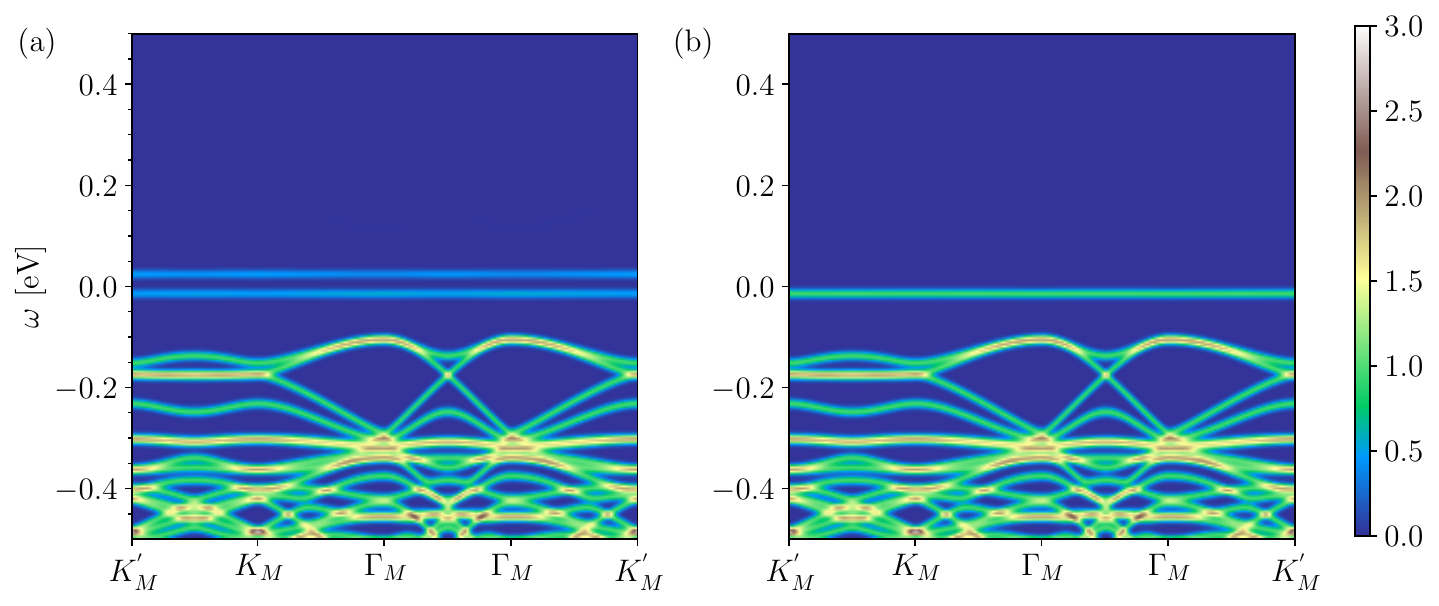}
    \caption{(Color online) Theoretical study of the time-resolved angle-resolved photoemission spectrometry (trARPES) spectrum of the periodically driven TBG along the path shown in Fig.~\ref{fig:occupation}(a). When the width of the probe pulse is much larger than the time period of the drive, i.e., \( \sigma_{\mathrm{probe}} \gg T \), trARPES matches closely to the occupation of Floquet bands shown in \cref{fig:occupation}. trARPES spectrum when (a) PR occupation, and (b) SS occupation is realized. The parameters used for the above plots are \(\Omega = 4.5 \)~eV and \(A_0 = 1.5k_{\theta}\),  corresponding to an electric-field magnitude of $E \approx \Omega A_0 = 2.2 \times 10^4$~kV/cm. The twist angle is $\theta=1.1^\circ$, and the chemical potential is $\mu = 0$. The width of the probe pulse is \( \sigma_{\mathrm{probe}} \approx 72 T\). }
    \label{fig:Arpes}
\end{figure*}
In this section, we compute the trARPES spectrum of the periodically driven TBG.
For analytical calculations we assume, \( \sigma_{\mathrm{pump}} \gg T \), 
and drop the Gaussian part from \( \bm A_{\mathrm{pump}}(t) \). In this limit, the vector potential for the pump pulse becomes periodic, and one can use tools from Floquet theory to simplify Eq.~\eqref{eq:occupation}.
The evolution operator in \cref{eq:evol-op}, in the Floquet basis is given by,
\begin{align} 
 \mathcal U_{\bm k}(t,0)= \sum_{\alpha} e^{-i \varepsilon_{\bm k \alpha} t }\ket{\phi_{\alpha}(\bm k , t)}\bra{\phi_{\alpha}(\bm k ,0)},
 \label{eq:ans}
\end{align} 
where \(\varepsilon_{\bm k \alpha}\) is $\alpha$th quasienergy band and \( |\phi_{\alpha}(\bm k,t)\rangle \) is the corresponding Floquet mode. Following Eq.~\eqref{eq:anh} and Eq.~\eqref{eq:ans}, one can simplify the trace of lesser Green's function in the Floquet basis as
\begin{align}
    \Tr[\mathcal G^{<}(\bm{k} ,t_1, t_2)] 
     &= \sum_{\alpha,\beta}\sum_{m,n} e^{i(\varepsilon_{\bm k \alpha} + m\Omega)t_{2}}
     e^{-i(\varepsilon_{\bm k \beta} + n\Omega)t_{1}}\nonumber \\
     &\times \langle \phi_{\beta}(\bm k,0)|\mathcal G^{<}(\bm k ,0,0) |\phi_{\alpha}(\bm k , 0)\rangle \nonumber \\
     &\times \langle \phi_{\alpha}^{(m)}(\bm k)|\phi_{\beta}^{(n)}(\bm k)\rangle.
 \label{eq:ank}
\end{align} 

For analytical simplicity, we assume \( t_{\mathrm{pr}} = t_{p} = 0 \) and \( t_{0} = NT \), where \( N \gg 1 \) is a large number. Substituting Eq.~\eqref{eq:ank} into Eq.~\eqref{eq:occupation} and performing the temporal integrals, one obtains the photoemission intensity $\mathcal I(\bm k, \omega)$ as:
\begin{widetext}
\begin{align}
	\mathcal{I}(\bm k,\omega) &= \frac{\pi \sigma_{\mathrm{probe}}^2}{4N^2 T^2}\sum_{\mathclap{\substack{\alpha,\beta \\ m,n}}}\Im\left[\matrixel{\phi_{\alpha}(\bm k , 0)}{\mathcal G^<(\bm k,0,0)}{\phi_{\beta}(\bm k ,0)}\langle{\phi^{(m)}_{\beta}(\bm k )}|{\phi^{(n)}_{\alpha}(\bm k )}\rangle\right] \nn \\
	&\times\exp(- \sigma^2_{\mathrm{probe}}(\omega - \varepsilon_{\bm k\alpha} - n \Omega)^2 / 2)
	\exp(- \sigma^2_{\mathrm{probe}}(\omega - \varepsilon_{\bm k\beta} - m \Omega)^2 / 2) \nn \\
	&\times \overline\erf\left(\frac{NT}{\sqrt 2 \sigma_{\mathrm{probe}}},\frac{i \sigma_{\mathrm{probe}}(\varepsilon_{\bm k\alpha} - \omega + n \Omega)}{\sqrt2}\right)
	\overline\erf\left(\frac{NT}{\sqrt 2 \sigma_{\mathrm{probe}}},\frac{i \sigma_{\mathrm{probe}}(\varepsilon_{\bm k\beta} - \omega + m \Omega)}{\sqrt2}\right),
  \label{eq:arp1}
\end{align}
\end{widetext}
where \( \overline{\mathrm{erf}}(a,b) = [\erf(a+b)+\erf(a-b)]/\sqrt{2} \) and $\erf(x)=2 \int_0^x \dd t \exp(-t^2) /\sqrt{\pi}$ is the error function. 
The energy resolution of the trARPES spectrum is inversely proportional to \(\sigma_{\mathrm{probe}}\).
The photoemission intensity \( \mathcal{I}(\bm{k}, \omega) \) of the occupied band depends on the overlap of the Fourier components of the Floquet mode, given by \( \langle \phi^{(m)}_{\beta}(\bm{k}) | \phi^{(n)}_{\alpha}(\bm{k}) \rangle \). This intensity is maximized when \( m = n \) and \( \alpha = \beta \), corresponding to \( \omega = \varepsilon_{\bm{k}\alpha} + n \Omega \). Otherwise, it is suppressed.
The expression simplifies further if we make the physically motivated assumption that the probing interval is much larger than the width of the probe pulse, i.e., \(NT \gg \sigma_{\mathrm{probe}}\). Under this assumption, the error function approximately becomes equal to 1, resulting in:
\begin{widetext}
\begin{align}
\mathcal I(\bm k, \omega)
\approx \frac{\pi \sigma^2_{\mr{probe}}}{8 N^2 T^2}
\sum_{\mathclap{\substack{\alpha,n}}}\Im\left[\matrixel{\phi_{\alpha}(\bm k , 0)}{\mathcal G^<(\bm k,0,0)}{\phi_{\alpha}(\bm k ,0)}\langle{\phi^{(n)}_{\alpha}(\bm k )}|{\phi^{(n)}_{\alpha}(\bm k )}\rangle\right]
	\exp(- \sigma^2_{\mathrm{probe}}(\omega - \varepsilon_{\bm k\alpha} - n \Omega)^2).
\end{align}
\end{widetext}

Depending on whether SS or PR occupation is realized in the system, \(\mathcal G^{<}(\bm{k}, 0, 0)\) has distinct forms.  For the theoretical calculation of the trARPES spectrum with SS occupation, we assume that a pump pulse was applied in the distant past to a system weakly coupled to a bath. Subsequently, the system achieves a steady-state at timescales longer than that of the system-bath coupling. In this case, the lesser Green's function is diagonal in the Floquet basis at a reference time \(t = t' = 0\), i.e., \(\mathcal{G}^{<}_{\alpha \beta}(\bm{k},0,0) = i n^{\mathrm{ss}}_{\alpha} \delta_{\alpha \beta}\), where \(n^{\mathrm{ss}}_{\alpha}\) is the steady-state occupation of the \(\alpha\)-th Floquet state, given by Eq.~\eqref{eq:steady-state-occu}. Fig.~\ref{fig:Arpes}(b) shows the trARPES spectrum of TBG, which closely resembles the steady-state occupation shown in Fig.~\ref{fig:occupation}(d).

For the case of PR occupation, assume that the system was in thermal equilibrium before the pump pulse was applied. Subsequently, the system is completely disconnected from the bath and becomes effectively closed. Thereafter, the pump pulse is applied to the system. In this case, the lesser Green's function at a reference time \(t = t' = 0\) becomes identical to the thermal density matrix  of the system $\rho^{\mr{static}}(\bm k)$ defined in \cref{eq:thermal-dens-mat}, i.e., 
\(\mathcal{G}^{<}(\bm{k}, 0, 0) = i \rho^{\mathrm{static}}(\bm{k})\).
Fig.~\ref{fig:Arpes}(a) shows the theoretical trARPES spectrum, that matches closely with the projected occupation shown in Fig.~\ref{fig:occupation}(c).

Additionally, we present the theoretical trARPES spectrum of Dirac and semi-Dirac systems showing PR and SS occupation in \cref{app:B}.

\color{black}

\begin{figure*}[ht]
    \centering
    \includegraphics[width=0.9\linewidth]{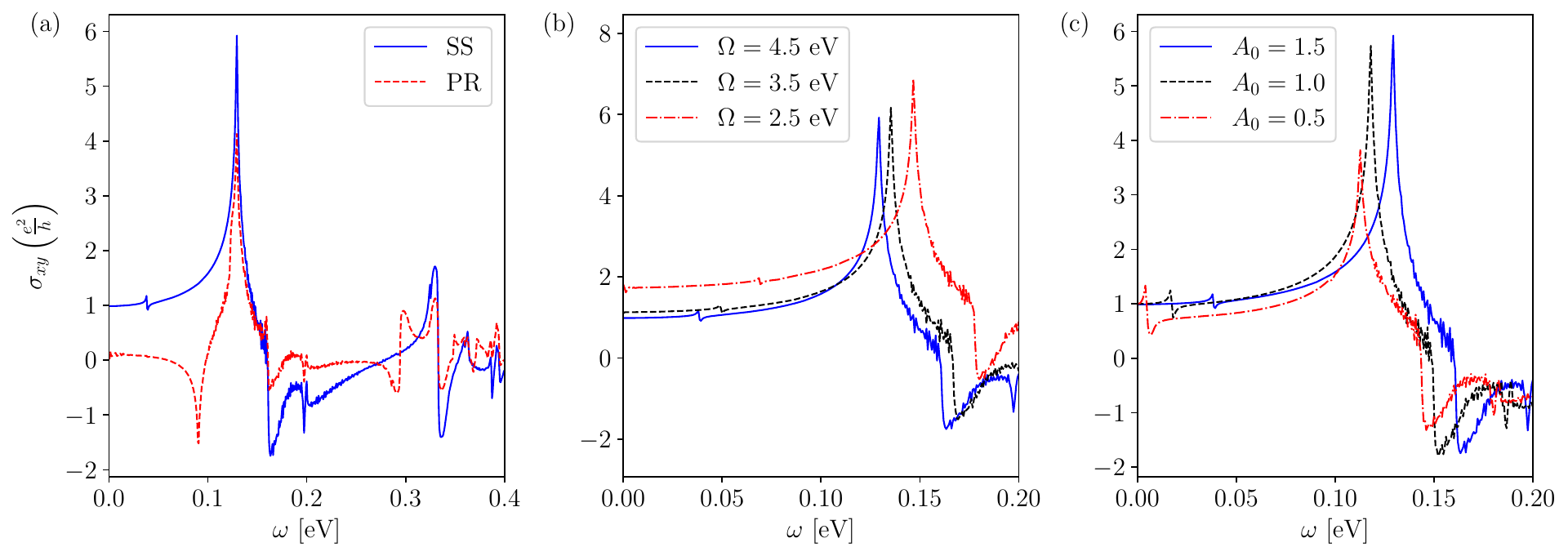}
    \caption{(a) Optical Hall conductivity $\sigma_{xy}$ of the TBG driven by circluar polarized light at the $K$-valley with PR (dashed line) and SS (solid line) occupation. The net optical Hall conductivity is four times (with valley and spin flavors) $\sigma_{xy}$ at valley $K$. The Optical Hall conductivity approaches quantized value with SS occupation. The parameters used for these plots are \(\Omega = 4.5\)~eV and \(A_{0} = 1.5 k_{\theta}\), \(\mu= 0\). (b) Optical Hall conductivity with SS occupation at different values of driving frequency and \(A_{0} = 1.5 k_{\theta}\). (c) Optical Hall conductivity with SS occupation at different values of driving amplitude (in the unit of \(k_{\theta}\)) and \(\Omega= 4.5\text{ eV}\). All driving parameters correspond to an electric field of the order of $10^4$~kV/cm.}
    \label{fig:cond}
\end{figure*}
\subsection{Optical Hall conductivity of periodically driven TBG}
\label{ssec:Optical Hall conductivity of periodically driven TBG}

In this section, we study the optical Hall conductivity of TBG driven with a circularly polarized light. We see several features in the optical conductivity that are different between the PR and SS occupations.
In a static insulating electronic system with broken time-reversal symmetry, one typically expects quantized DC Hall conductivity at zero temperature, where the Fermi-Dirac distribution governs the occupations of the bands. When the system is driven out of equilibrium, the occupancy of the band is no longer a Fermi-Dirac distribution, and quantization becomes less apparent. Nevertheless, in cases where the occupancy resembles that of a band insulator, the Hall conductivity tends towards quantization \cite{oka2009photovoltaic}. Our study confirms this assertion, particularly under large and moderate driving frequency with SS occupation.

 For the case of SS occupation, we observe a nearly quantized DC optical Hall conductivity (see Fig.~\ref{fig:cond}(a)) for a driving frequency \(\Omega = 4.5\) eV and vector potential amplitude \(A_0 = 1.5 k_{\theta}\). These drive parameters correspond to an electric field of the order of \(10^4\) kV/cm, which is achievable with current laser technology \cite{Li_2020}.
 In this case, we obtain a DC Hall conductivity of 
 $\approx 4 e^2 / h$ (accounting for both valleys and spins), that corresponds to Chern number of 4, as was previously reported in Ref. \cite{Li_2020}.
 This can be understood from the SS occupation Fig.~\ref{fig:occupation}(d), where the occupation of the flat conduction and valance bands in the chiral limit mirrors that of a band insulator. Circularly polarized light breaks the time-reversal symmetry, inducing mass terms with opposite signs in each valley of graphene. Consequently, the combined Chern numbers stemming from the valleys and spin-flavors in graphene sum up, yielding a total Chern number of $\pm 4$.

 On the other hand, this does not hold true for the PR occupation. In this case, both flat bands exhibit equal occupations ($\approx 1/2$ for each band), as depicted in Fig.~\ref{fig:occupation}(c). The optical Hall response differs markedly between SS and PR occupations. With SS occupation, we observe a kink at low energy, around 0.05 eV, stemming from optical transitions occurring between the split flat bands (with a gap of the order of $\sim\mathcal O(A_0^2/\Omega)$) induced by circularly polarized light. Conversely, such a kink is absent in the Hall response when PR occupation is realized, as optical transitions between the flat bands are Pauli blocked. 
 As we increase the frequency ($\omega$), we observe peaks in the optical Hall response attributed to optical transitions involving higher energy bands. In the case of SS occupation, the first peak in $\sigma_{xy}$ signifies transitions from the valence band to the second conduction band. Similarly, a comparable peak is noted at a slightly lower frequency for PR occupation, reflecting transitions between the first and second conduction bands of the driven TBG. Further increasing the energy reveals multiple peaks, indicative of transitions to additional higher energy bands.
 
 By changing the driving frequency, the system can undergo
 topological phase transitions. In Fig.~\ref{fig:cond}(b), we illustrate the $\Omega$ dependence
 of Hall conductivity with SS occupation for some representative values of $\Omega$. 
 Transitioning from high-frequency driving ($\Omega=4.5$~eV) to moderate driving frequencies ($\Omega=2.5$~eV), the system undergoes a topological phase transition from $\sigma_{xy} \approx 4~e^2/h$ to $\sigma_{xy} \approx 8~e^2 / h $. However, high-energy features remain qualitatively similar, albeit occurring at different $\omega$ values depending on $\Omega$. 
 
 In Fig.~\ref{fig:cond}(c), we show the dependence of optical Hall conductivity on the driving amplitude for \(\Omega = 4.5\)~eV. With increasing driving amplitude, the separation between flat bands increases, and corresponding kinks in the plot of optical Hall conductivity shift towards a higher value of \(\omega\). However, with all values of driving amplitude, we get a nearly quantized DC Hall conductivity for the SS occupation.

 Additionally, we present the optical Hall conductivity of Dirac and semi-Dirac systems with SS and PR occupation, as shown in \cref{app:B}. 

\section{Summary}
\label{sec:Summary}
We theoretically studied the trARPES spectrum and optical conductivity of driven twisted bilayer graphene (TBG), near the flat-band limit. The periodic drive can break time-reversal symmetry, giving rise to net Chern numbers of quasienergy bands. If the periodic drive, which we call the pump field, is on for a relatively long time, such that the system is in a pre-thermal steady-state, one assumes a steady-state (SS) occupation of the quasienergy bands, then the optical conductivity gives rise to near-quantized signatures, reminiscent of non-trivial band topology. On the other hand, when the response is not from a steady-state, the quasienergy bands follow a projected (PR) occupation, and the responses differ from quantized values. These two possible occupations of the quasienergy bands have contrasting responses in the trARPES spectrum as well. Apart from TBG, we compute the same in Dirac as well as semi-Dirac semi-metallic systems, pointing out differences in response for PR and SS-occupied states. 
In the later case, we assume that the system is already in a steady-state, without addressing the timescale required to reach this state, as it may depend on the driving protocol and specific details of the system, making it non-universal.
At high driving frequencies, the steady-state occupation is predominantly determined by the zeroth harmonic of the periodic part of the Floquet state, closely resembling the ideal occupation at zero temperature. In contrast, at low driving frequencies, the Floquet states undergo significant deformation, with multiple harmonics of the periodic component contributing to the occupation. In this regime, the occupation deviates from the ideal case, and observables associated with such an occupation no longer have direct counterparts in the static system.
Our findings are applicable to any periodically driven electronic systems and can be validated through various existing experimental techniques. The occupation of the Floquet bands can be observed in pump-probe experiments \cite{wang2013observation}. Additionally, the DC limit of optical Hall conductivity can be measured in multiterminal experiments \cite{torres2014multiterminal} as well as in time-of-flight measurements in optical lattices \cite{jotzu2014experimental}.

\begin{acknowledgments}
RK acknowledges funding under the PMRF scheme (Govt. of India). AK acknowledges support from the SERB (Government of India) via Sanction No. ECR/2018/001443 and CRG/2020/001803, DAE (Government of India) via Sanction No. 58/20/15/2019-BRNS, and MHRD (Government of India) via Sanction No. SPARC/2018-2019/P538/SL.
\end{acknowledgments}
\bibliography{refs}
\onecolumngrid
\appendix
\counterwithout{equation}{section}
\setcounter{equation}{0}
\renewcommand{\theequation}{S\arabic{equation}}
\setcounter{figure}{0}
\renewcommand{\thefigure}{S\arabic{figure}}

\section{Optical conductivity of periodically driven TBG}
\label{app:A}
According to the linear response theory, the real part of optical conductivity is related to the imaginary part of the current-current correlation. The retarder current-current correlation function is given by \cite{Dehghani_2015},
\begin{equation}
    \chi_{ij}(\bm q,t,t')= -i\theta(t- t')\langle[J_{\bm qI}^{i}(t),J_{- \bm q I}^{j}(t')]\rangle,
     \label{eq:b1}
\end{equation}
where \(J_{\bm q I}^{i}(t)\) is current operator in direction \(\hat{i}\) in interaction picture. 
In the above expression \(\langle\cdot\rangle \equiv \mathrm{Tr}(\hat{\rho}\cdot)\), where \(\hat{\rho}\) is the density matrix of the system.
The current operator in the direction \(\hat{i}\) in the interaction picture is given by,
\begin{equation}
    J_{\bm q I}^{i}(t)= \frac{1}{\sqrt{V}}\sum_{\bm k}c_{\bm k + \frac{\bm q}{2}}^{\dagger}(t)\frac{\partial \mathcal H(\bm k, t)}{\partial k_{i}}c_{\bm k -\frac{\bm q}{2} }(t).
     \label{eq:b2}
\end{equation}
Here, $V$ is the total area of the system. 
The correlation function in the momentum-frequency space reads \cite{Kumar_2020}: 
 \begin{equation}
\begin{aligned}
  \chi_{ij}(\bm q,\omega) &= \frac{1}{V}\sum_{\bm k}\sum_{\alpha \beta}\sum_{n}\frac{D^{(n)}_{\alpha i \beta}\left(\bm k + \frac{\bm q}{2}\right)D^{(-n)}_{\beta j \alpha}\left(\bm k - \frac{\bm q}{2}\right)(n_{\bm k + \frac{\bm q}{2}\alpha} - n_{\bm k - \frac{\bm q}{2}\beta})}{\omega + i\delta + \varepsilon_{\bm k + \frac{\bm q}{2} \alpha}- \varepsilon_{\bm k + \frac{\bm q}{2} \beta}- n \Omega},
     \label{eq:b11}
\end{aligned}
\end{equation}
where \( D^{(n)}_{\alpha i \beta}(\bm k )= \sum_{p, q}\matrixel{\phi^{(p)}_{\bm k \alpha}}{\frac{\partial \mathcal H^{(n+ p-q)}(\bm k)}{\partial k_{i}}}{\phi^{(q)}_{\bm k \beta}}\), and \(n_{\bm k \alpha}\) is the occupation of the \(\alpha\)th Floquet state. In this formula \(n_{\bm k \alpha}\) represents either the projected (Eq.~\eqref{eq:quasi-steady-state-occu}) or steady-state occupation (Eq.~\eqref{eq:steady-state-occu}).
If the quasi-energy bands are gapped at the chemical potential, the above equation in the thermodynamic limit $\bm q \to 0$ reduces to,
 \begin{equation}
\begin{aligned}
  \chi_{ij}(\omega) &= \frac{1}{N}\sum_{\bm k}\sum_{\alpha \beta}\sum_{n}\frac{D^{(n)}_{\alpha i \beta}\left(\bm k \right)D^{(-n)}_{\beta j \alpha}\left(\bm k \right)(n_{\bm k \alpha} - n_{\bm k \beta})}{\omega + i\delta + \varepsilon_{\bm k \alpha}- \varepsilon_{\bm k \beta}- n \Omega}.
     \label{eq:b111}
\end{aligned}
\end{equation}
The real part of optical Hall conductivity is then given by,
\begin{equation}
    \Re \sigma_{xy}(\omega)= -\frac{\Im \chi_{xy}(\omega)}{\omega}.
\end{equation}
\label{app: Optical
conductivity of periodically driven TBG}
\begin{figure*}[ht]
    \centering
    \includegraphics[width=246pt]{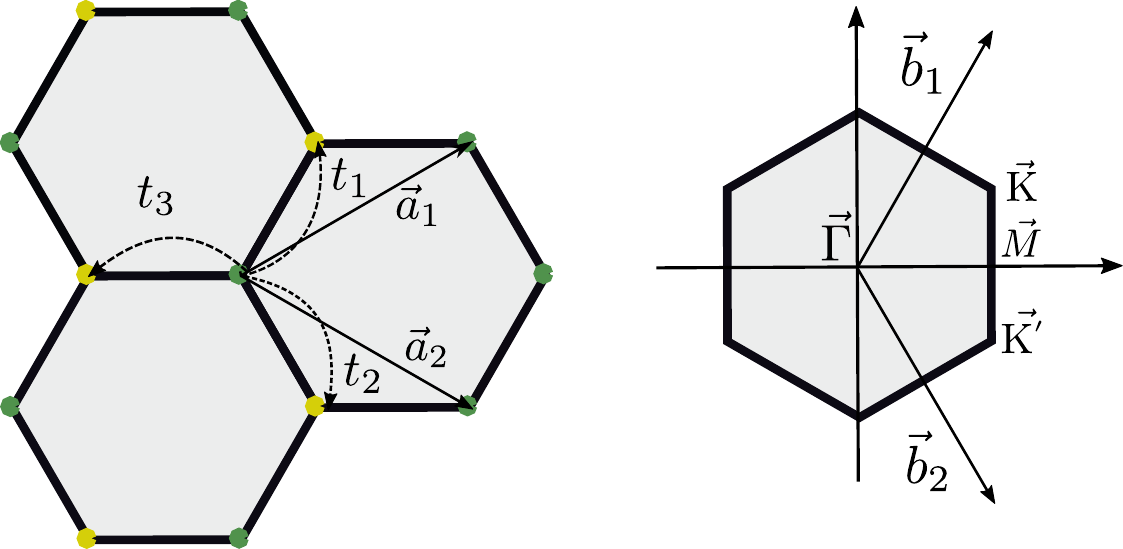}
    \caption{The honeycomb lattice (left) and the first Brillouin zone of the monolayer graphene (right). The primitive reciprocal lattice vectors are given by \(\bm b_{1}= \frac{2\pi}{3a}\left(1, \sqrt{3}\right)\), \(\bm b_{2}= \frac{2\pi}{3a}\left(1, -\sqrt{3}\right)\).}
    \label{fig:lattice}
\end{figure*}
\section{Additional results for trARPES and optical Hall conductivity of Dirac and semi-Dirac semimetal}
\label{app:B}
We consider a tight-binding model for the honeycomb lattice with spatially anisotropic hopping.
The Bloch Hamiltonian for the tight binding model is given by,
\begin{align}
    \mathcal H(\bm k)= \begin{pmatrix}
		0 & g(\bm k) \\
		g(\bm k)^{*} & 0
	\end{pmatrix},
\end{align}
where \(g(\bm k)= t_{1}e^{i\bm k \cdot \bm a_{1}} + t_{2}e^{i\bm k \cdot \bm a_2} + t_{3}\) with \(t_{i}\) being the anisotropic hopping amplitudes to nearest neighbors, as shown in the Fig.~\ref{fig:lattice}. Here, the primitive lattice vectors of the hexagonal lattice are given by: \(\bm a_{1}= a\left(\frac{3}{2}, \frac{\sqrt{3}}{2}\right)\), \(\bm a_{2}= a\left(\frac{3}{2}, -\frac{\sqrt{3}}{2}\right)\) with $a$ being the lattice constant.
We drive the system using a circularly polarized light. The vector potential for the same is given by \(\bm A(t)= A_{0}(\cos(\Omega t), \sin(\Omega t))\). In the presence of circularly polarized light, the time-dependent Bloch Hamiltonian is given by, 
\begin{align}
\mathcal H(\bm k, t) = 
     \begin{pmatrix}
		0 & g(\bm k + \bm A(t)) \\
		g(\bm k + \bm A(t))^{*} & 0
	\end{pmatrix},
\end{align}
where
\begin{align}
 g(\bm k + \bm A(t))= t_{1}e^{i\bm k \cdot \bm a_{1} + i\bm A(t) \cdot \bm \delta_{1}} + t_{2}e^{i\bm k \cdot \bm a_{2} + i\bm A(t) \cdot \bm \delta_{2}} + t_{3}e^{ i\bm A(t) \cdot \bm \delta_{3}},
  \label{eq:fab}
\end{align}
with \(\bm \delta_{1}= a\left(\frac{1}{2}, \frac{\sqrt{3}}{2}\right)\), \(\bm \delta_{2}= a\left(\frac{1}{2}, -\frac{\sqrt{3}}{2}\right)\), \(\bm \delta_{3}= a\left(-1,0\right)\). 
 The Fourier components of the Hamiltonian reads,
 \begin{align}
    \mathcal{H}^{(n)}(\bm k)= \begin{pmatrix}
            0 & g^{(n)}(\bm k)\\
            g^{(n)}(\bm k)^*&0
           \end{pmatrix},
\end{align}
 where $g^{(n)}(\bm k)$ is given by,
\begin{align}
  g^{(n)}(\bm k) &= \frac{1}{T}\int_{0}^{T}e^{i n \Omega t} g(\bm k + \bm A(t) ) \dd t \nonumber \\
  &= t_{1}e^{i(\frac{3 kx}{2}+ \frac{\sqrt{3} ky}{2})}e^{i n \frac{\pi}{6}}\mathcal{J}_{n}(A_{0}) + t_{2}e^{i(\frac{3 kx}{2}- \frac{\sqrt{3} ky}{2})}e^{-i n \frac{\pi}{6}}(-1)^{n}\mathcal{J}_{n}(A_{0}) + t_{3}i^{n}(-1)^{n}\mathcal{J}_{n}(A_{0}),
  \label{eq:fcomp}
\end{align}
where \(\mathcal{J}_n(.)\) is the \(n\)th Bessel function of first kind. For numerical calculations, we choose: i)  \(t_{1} = t_{2} = t_{3}= \lambda\) for Dirac and ii) \(t_{1} = t_{2} = \lambda,~ t_{3}= 2\lambda\) for a semi-Dirac semimetal. We set the energy scale as $\lambda$ and the length scale as $a$. Here, we discuss the steady-state (SS) occupation and projected (PR) occupation of the Dirac and semi-Dirac models. Subsequently, we present the theoretical trARPES spectrum and optical Hall conductivity, showing the signature of either occupation for each of the model.

\begin{figure*}[ht]
    \centering
    \includegraphics[width=0.9\linewidth]{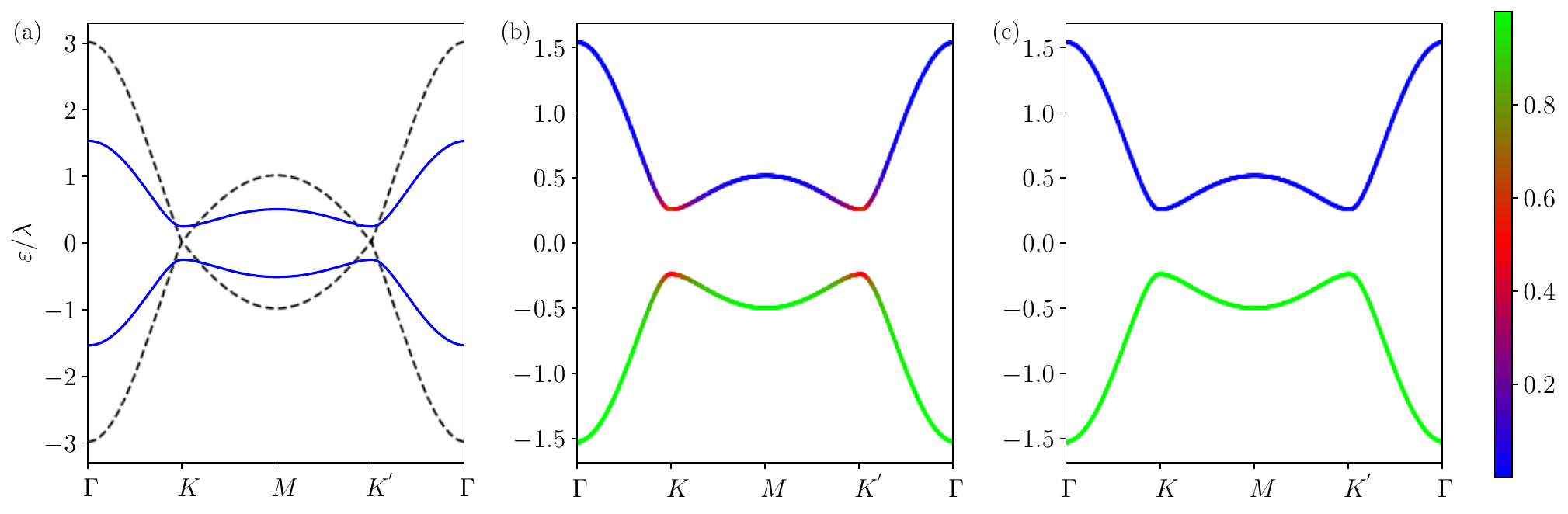}
    \caption{Band structure and occupation of the periodically driven single layer graphene (Dirac-system) along the high symmetry points shown in Fig.~\ref{fig:lattice}. (a) Static band structure (dashed line) and quasi-energy band structure (solid line). 
    The bands are colored by the value of
    PR and SS occupation in (b) and (c), respectively.
    The parameters used in this plot are \(\Omega = 10\lambda\), \(A_{0} = 1.5\).}
    \label{fig:band-dirac}
\end{figure*}

\begin{figure*}[ht]
    \centering
    \includegraphics[width=0.9\linewidth]{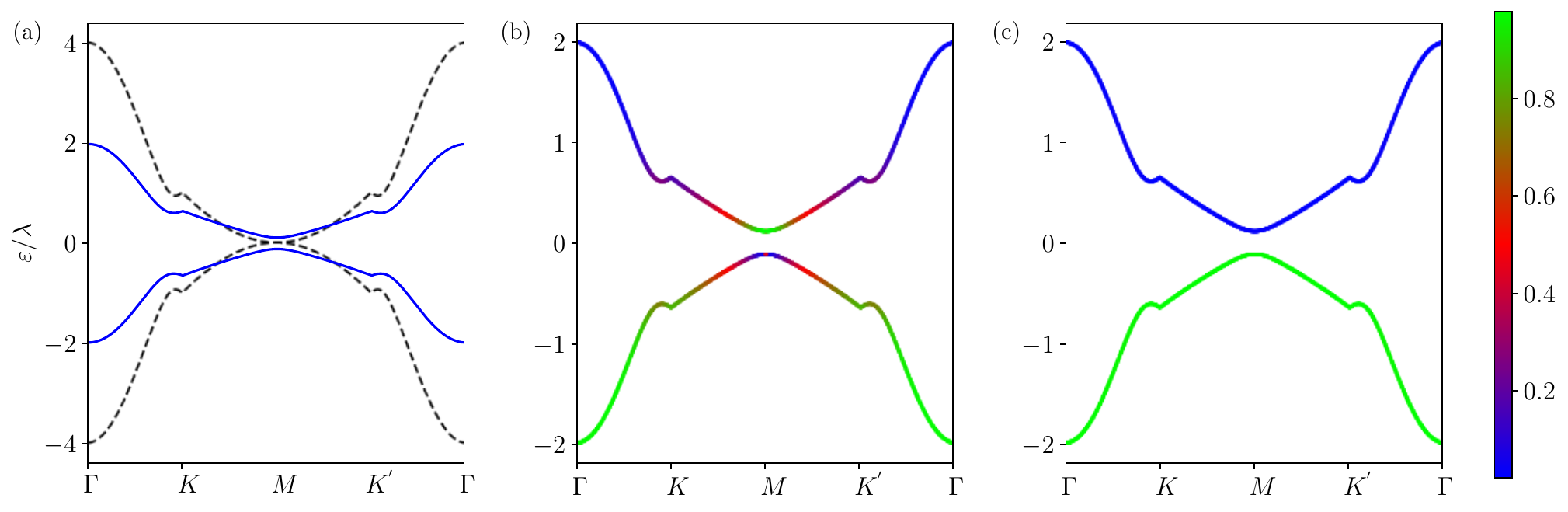}
    \caption{Band structure and occupation of the periodically driven single layer graphene (semi-Dirac system) along the high symmetry points shown in Fig.~\ref{fig:lattice}. (a) Static band structure (dashed line) and quasi-energy band structure (solid line). (b) projected (PR) occupation, which does not incorporate any relaxation mechanism. (C) steady-state (SS) occupation, which takes into account the relaxation mechanism because of the coupling between the system and the bath. The parameters used in this plot are \(\Omega = 8\lambda\), \(A_{0} = 1.5\).}
    \label{fig:band-semi-dirac}
\end{figure*}

The static band structure of the tight binding Dirac model is gapless at high symmetry points $K$ and \(K^{'}\) whereas the quasienergy band structure shows gaps at these points, as shown in Fig.~\ref{fig:band-dirac}(a). The opening of gaps in the quasienergy band structure is due to the broken time-reversal symmetry of the system by periodic drive (circularly polarized light). This leads to the redistribution of occupation in the quasienergy bands around these gaps. Fig.~\ref{fig:band-dirac}(b) and Fig.~\ref{fig:band-dirac}(c) show the PR and SS occupation for the Dirac model, respectively. One can easily note that the SS occupation is nearly one and zero for the quasienergy band below and above the gap, respectively, whereas the PR occupation is nearly half for either band around the gap. These generic features are also true for the semi-Dirac model as well (see Fig.~\ref{fig:band-semi-dirac}). 

\begin{figure*}[ht]
    \centering
    \includegraphics[width=0.8\linewidth]{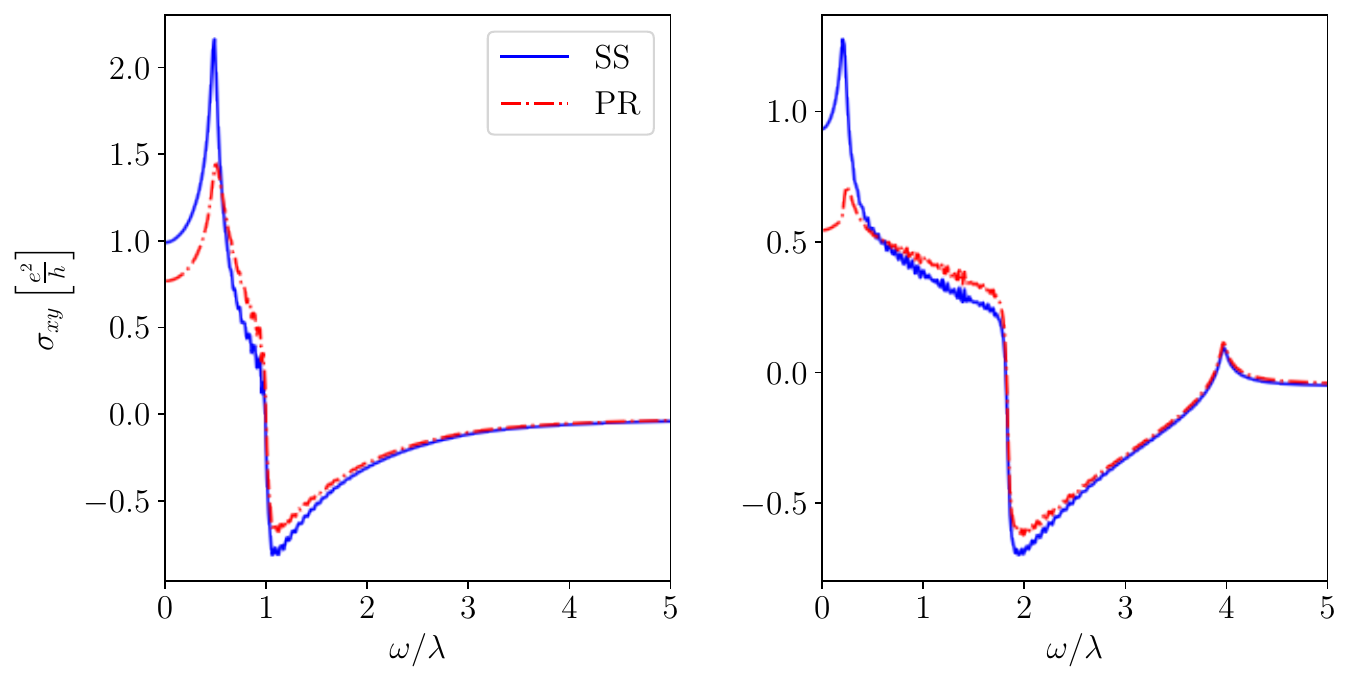}
    \caption{Optical Hall conductivity ($\sigma_{xy}$) of the Dirac Hamiltonian with SS (solid line) and PR occupation (dashed line).  
    The left figure shows $\sigma_{xy}$ for the Dirac model. The right figure is for the semi-Dirac model. The parameters used in this plot are the same as the ones used in Fig.~\ref{fig:band-dirac} and Fig.~\ref{fig:band-semi-dirac} for Dirac and semi-Dirac, respectively.}
    \label{fig:band-dirac-cond}
\end{figure*}

\begin{figure*}
   \centering
    \includegraphics[width=0.8\linewidth]{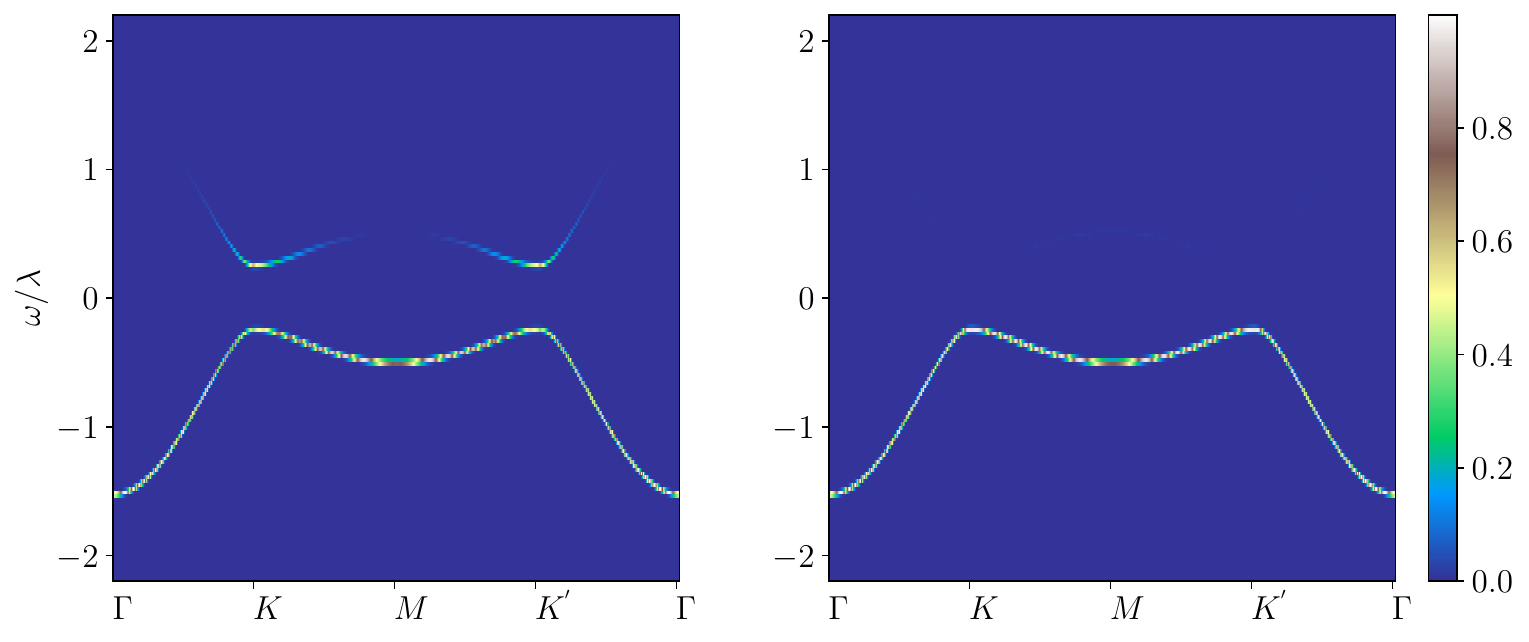} 
    \caption{trARPES of the Dirac model for
    projected occupation (left) and steady-state occupation (right). The value of \(\sigma_{\mathrm{probe}} \approx 95T\). The rest of the parameters used are the same as in Fig.~\ref{fig:band-dirac}.}
    \label{fig:arp1}
\end{figure*}

\begin{figure*}
   \centering
    \includegraphics[width=0.8\linewidth]{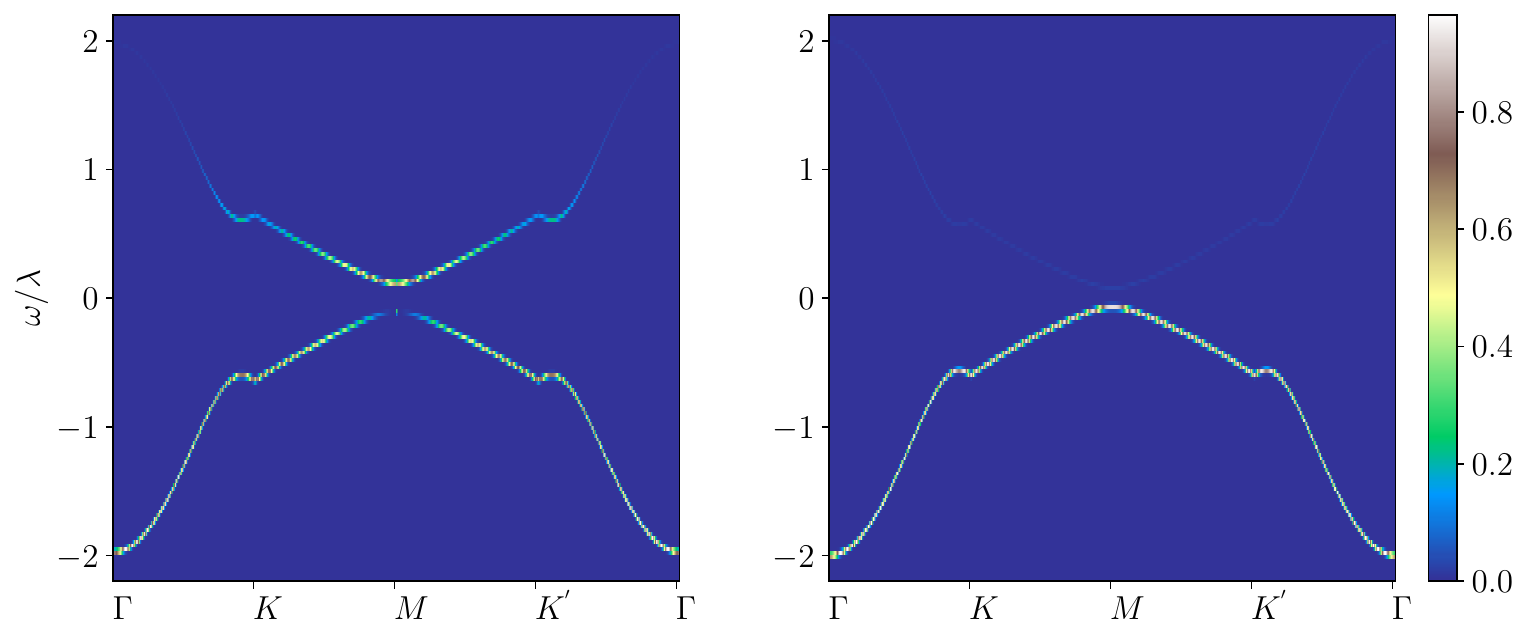} 
    \caption{trARPES of semi-Dirac model with
    projected occupation (left) and steady-state occupation (right).The value of \(\sigma_{\mathrm{probe}} \approx 76T\). The rest of the parameters used are the same as in Fig.~\ref{fig:band-semi-dirac}.}
     \label{fig:arp2}
\end{figure*}

For both of the models, we present the trARPES spectrum when PR or SS occupation is realized. Fig.~\ref{fig:arp1} and \ref{fig:arp2} show the trARPES spectrum, illustrating the influence of PR and SS occupation for the Dirac and semi-Dirac models, respectively. For both of the models, the DC optical Hall conductivity is nearly quantized with SS occupation, whereas it is not quantized with PR occupation, as shown in Fig. \ref{fig:band-dirac-cond}. The kinks appearing at higher energies are indicative of interband transitions. For more details, refer to \cref{Sec:results} of the main text.    

\section{Degenrate perturbation theory to compute the value of steady-state (SS) and projected (PR) occupation in the quasienergy bands around gaps}
\label{app:C}
In this section, we apply the degenerate perturbation theory to compute the value of SS and PR occupations of the quasienergy band around the drive-induced band gaps. Such gap opening can be of two kinds. If the static bands have degeneracies protected by symmetry, then the driven can break symmetry and lift the degeneracies. An example of this kind is the topological gap opening at the Dirac point of irradiated graphene with circularly polarized light at high frequency. At zero temperature, we show below that if such gap opening takes place at the chemical potential, the lower quasienergy band remain occupied leaving the upper quasienergy band unoccupied.

At relatively lower frequency, the static bands go through `band-folding', and folds in the Floquet zone to form quasienergies. As the bands fold on themselves, there are degeneracies and opening of gaps at these points. At zero temperature, we show that whenever such gap opening takes place by mixing an occupied and an unoccupied band of the static system, it results in an equal distribution of the occupation at the point of gap opening.

\subsubsection*{Gap opening by breaking symmetry}
Consider a degenerate eigenvalue (which we also take to be at the chemical potential, $\mu = 0$) with eigenstates \(|\psi_{\pm}\rangle\) at some momentum point. If the driving frequency is high with a small amplitude, the effective Hamiltonian at this point is given by high-frequency approximation as~\cite{kitagawa2011transport}
\begin{equation}
\begin{aligned}
\mathcal{H}_{\mathrm{eff}} = \mathcal{H}_{0} + \delta \mathcal{H}+ \ldots,
\end{aligned}
\label{eq:d20}
\end{equation}
where $\mathcal{H}_{0}$ is the static part of the Hamiltonian with a symmetry that protects the degeneracy. \(\delta \mathcal{H}= \left[\mathcal{H}^{(+1)},\mathcal{H}^{(-1)}\right]/{\Omega}\) can be treated as a perturbation where we neglect higher-order terms in $1/\Omega$. If the degeneracy is lifted by this perturbation, at high frequency, the corrected energies, \( \epsilon_{\pm} = |\langle \psi_{-}| \delta \mathcal{H}| \psi_{+}\rangle |\), are also the quasienergies. The corresponding eigenstates are the zeroth Fourier components of the corresponding Floquet states. Further, for weak driving, non-zero Fourier components of these Floquet states can also be neglected as they are expected to only grow polynomially with the driving amplitude~\cite{kitagawa2011transport}.

The resulting Floquet states can thus be written as $|\phi_{\pm}(t)\rangle \approx \left(|\psi_{+}\rangle+e^{i\theta}|\psi_-\rangle\right)/\sqrt{2}$, which is static in the limit of small driving amplitude \cite{Biswas_2020}. Here $\theta = \text{arg}(\langle \psi_{-}| \delta \mathcal{H}| \psi_{+}\rangle)$.  The SS occupation of these states are $ n_{\pm}^{\rm ss} = \sum_{l}f(\varepsilon_{\pm}+ l\Omega)\langle\phi_{\pm}^{(l)}|\phi_{\pm}^{(l)}\rangle$. As only the \(l =0\) Fourier component survives, at zero temperature \(n_{-}^{\rm ss} = 1\) and \(n_{+}^{\rm ss}= 0\) as \(f(\varepsilon_{-}) = 1\) and \(f(\varepsilon_{+}) = 0\).

The density matrix before driving can be written as $\rho^{\mathrm{static}} = \frac12\left(|\psi_-\rangle\langle \psi_-| + |\psi_+\rangle\langle \psi_+|\right)$. It is easy to deduce that the PR occupations, $n_{\pm}^{\rm pr}= \langle\phi_{\pm}|\rho^{\mathrm{static}}|\phi_{\pm}\rangle = 1/2.$ 
\subsubsection*{Gap opening at band-folded degeneracies}
At lower frequencies, there can be degeneracies as the folded quasienergy bands overlap within the Floquet zone. Let us consider that such overlaps are between folded static bands that were originally occupied and unoccupied. In the absence of driving, an extended zone picture (see Eq.~\eqref{eq:ex-ham} of main text) of a static Hamiltonian consists of only diagonal blocks. This leads to band-folding within a Floquet zone $(-\Omega/2,\Omega/2]$ with level crossings. If $E_{\pm}$ are the static energies of a system with $E_{-}$ and $E_{+}$ being occupied and unoccupied, respectively, then the band-foldings results in a degeneracy at the quasienergy $\varepsilon$ if $E_{+}+p\Omega = E_{-}+q\Omega = \varepsilon$, where $\varepsilon\in (-\Omega/2,\Omega/2]$, for integers $p$ and $q$. The corresponding Floquet states are $|\phi_+(t)\rangle=e^{ip\Omega t}|\psi_+\rangle$ and $|\phi_-(t)\rangle=e^{iq\Omega t}|\psi_-\rangle$, i.e., they have only $(-p)$th and $(-q)$th Fourier components. In the extended-zone basis, the eigenstates are large column vectors with only $(-p)$the and $(-q)$th row being non-zero, with elements of $|\psi_+\rangle$ and $|\psi_-\rangle$, respectively. Let us represent these large column vectors by $|\Phi_+\rangle\rangle$ and $|\Phi_-\rangle\rangle$.

In the presence of driving, the extended-zone Hamiltonian is modified with off-diagonal terms, which we denote as $\mathcal{H}^{O}$. Such off-diagonal elements lead to avoided crossings at degeneracies. Degenerate perturbation theory at eigenvalues $\varepsilon$ of the extended-zone Hamiltonian yields correction to the eigenvalues
\begin{equation}
\begin{aligned}
\Tilde{\varepsilon}_{\pm} = \varepsilon \pm |z|,
\label{eq:d4}
\end{aligned}
\end{equation}
where $z = \langle\langle \Phi_+| \mathcal{H}^{O} |\Phi_+\rangle\rangle$. The modified eigenstates in the extended-zone picture are given by
\begin{equation}
\begin{aligned}
|\Tilde{\Phi}_{-} \rangle\rangle &= -\frac{e^{i\theta}}{\sqrt{2}}|\Phi_- \rangle\rangle + \frac{1}{\sqrt{2}}|\Phi_{+}\rangle\rangle, ~~|\Tilde{\Phi}_{+} \rangle\rangle &= \frac{e^{i\theta}}{\sqrt{2}}|\Phi_{-} \rangle\rangle + \frac{1}{\sqrt{2}}|\Phi_{+} \rangle\rangle,
\end{aligned}
\end{equation}
where \(\theta =  \mathrm{arg}(z)\). In terms of the time-dependent representation, the Floquet states are given by
\begin{equation}
\begin{aligned}
|\Tilde{\phi}_{-}(t) \rangle &= -\frac{e^{i\theta + iq\Omega t}}{\sqrt{2}}|\psi_- \rangle + \frac{e^{ip\Omega t}}{\sqrt{2}}|\psi_{+}\rangle, ~~|\Tilde{\phi}_{+}(t) \rangle &= \frac{e^{i\theta + iq\Omega t}}{\sqrt{2}}|\psi_{-} \rangle+ \frac{e^{ip\Omega t}}{\sqrt{2}}|\psi_{+} \rangle.
\label{eq:flpert}
\end{aligned}
\end{equation}

\begin{figure*}[t]
    \centering
    \includegraphics[width=1.0\linewidth]{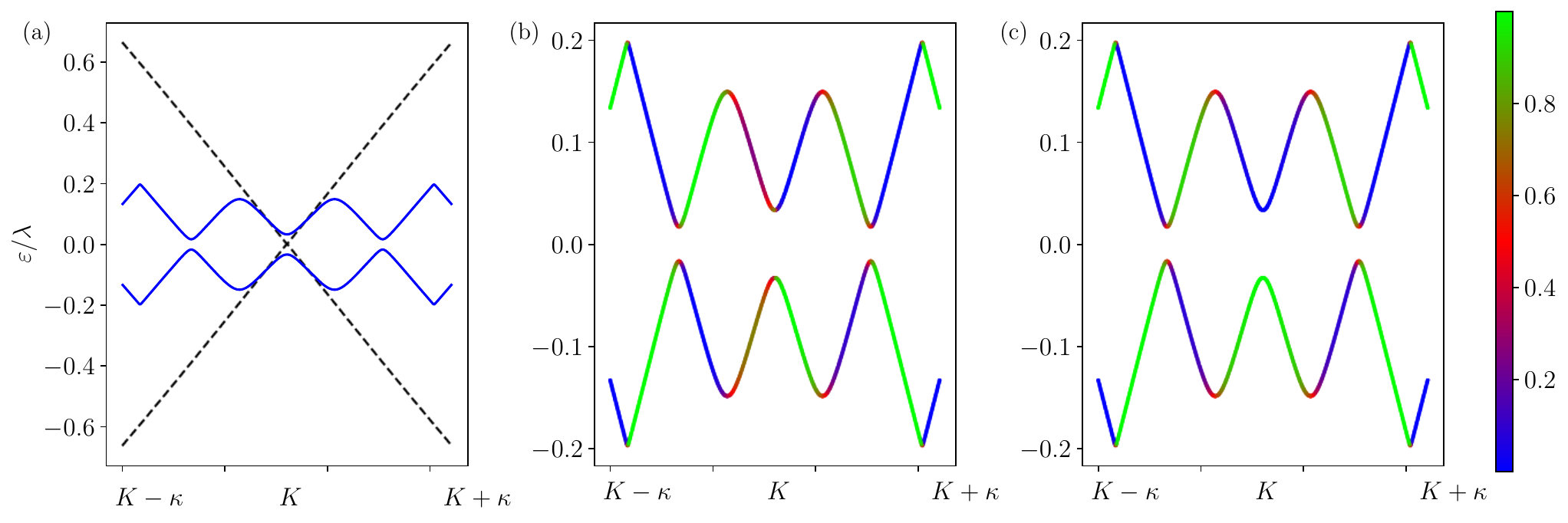}
    \caption{Band structure and occupation of the periodically driven single layer graphene (Dirac-system) around high symmetry point K. (a) Static band structure (dashed line) and quasi-energy band structure (solid line). (b) Projected (PR) occupation (c) steady-state (SS) of the Dirac model. The parameters used in for these plots are \(\Omega = 0.4\lambda\), \(A_{0} = 0.08\), \(\kappa = (0.45,0.0)\). At Dirac point K, the value of SS occupation is nearly one and zero for the lower and upper bands, respectively, at \(\varepsilon = 0\). In contrast, the value of PR occupation is nearly half for each band. Additionally, the value of SS and PR occupation is half for the quasienergy bands around a gap at \(\varepsilon =0\) (other than one at $K$) and \(\varepsilon =\frac{\Omega}{2}\).
} 
    \label{fig:dirac-occ-spliting}
\end{figure*}
The steady-state (SS) occupation of these states are:
\begin{equation}
\begin{aligned}
n_{\pm}^{\rm ss} = \sum_{l}f(\Tilde{\varepsilon}_{\pm}+ l\Omega)\langle\Tilde{\phi}_{\pm}^{(l)}|\Tilde{\phi}_{\pm}^{(l)}\rangle.
\end{aligned}
\end{equation}
In the above summation only the \((-q)\) and \((-p)\) Fourier components contribute, hence
\begin{equation}
\begin{aligned}
n_{+}^{\rm ss} &= f(\Tilde{\varepsilon}_{+}-p\Omega )\langle\Tilde{\phi}_{+}^{(-p)}|\Tilde{\phi}_{+}^{(-p)}\rangle + f(\Tilde{\varepsilon}_{+}-q\Omega )\langle\Tilde{\phi}_{+}^{(-q)}|\Tilde{\phi}_{+}^{(-q)}\rangle ,\\
&= \frac{1}{2}f(E_{+}+ |z| )+ \frac{1}{2}f(E_{+}+ |z| + (p-q)\Omega ).
\end{aligned}
\end{equation}
As $E_+ - E_- = (q-p)\Omega$, we obtain
$$n_{+}^{\rm ss} = \frac{1}{2}f(E_{+}+ |z| )+ \frac{1}{2}f(E_{-}+ |z|).$$
As $|z|$ is small, one can assume that at zero temperature, $f(E_+ + |z|) = 0$, and $f(E_- + |z|) = 1$, yielding $n_+^{\rm ss} = 1/2.$ A similar analysis also shows $n_-^{\rm ss} = 1/2.$

Considering $\rho^{\rm static} = |\psi_-\rangle\langle\psi_-|$, it is easy to arrive at $n^{\rm pr}_{\pm} = \langle \Tilde{\phi}_{\pm}(0)|\rho^{\rm static} |\Tilde{\phi}_{\pm}(0)\rangle = 1/2$.\\

Our analysis is agnostic to the details of any model. In Fig.~\ref{fig:dirac-occ-spliting}, we have shown the SS and PR occupations of the quasienergy bands when the system is driven with a low-frequency circularly polarized light. We see that around gaps \(\varepsilon =0, \frac{\Omega}{2}\) for the Dirac model (around Dirac point K), which is consistent with our calculation.  
We also note that at very low driving frequencies (as shown in Fig.~\ref{fig:dirac-occ-spliting}) the quasienergies are modified drastically. In such cases both occupations may mimic band inversion away from gapless points.

The overlap in the projected occupation, consisting of \(|\psi_n(\bm k)\rangle\) and \(|\phi_{\alpha}(\bm k, 0)\rangle\), exhibits different symmetry properties.
\(|\psi_n(\bm k)\rangle\) obeys \(\mathcal{P} \mathcal{T}\) symmetry, where \(\mathcal{P}\) is the inversion operator and \(\mathcal{T}\) is the time-reversal operator. This symmetry is crucial for protecting the gapless point. However, the periodic drive (circular polarized light) breaks this symmetry, resulting in the opening of a gap.
As a result, the symmetry of the occupation of bands under the transformation \(\bm k \leftrightarrow -\bm k\) near the K and K$'$ points has been broken, as shown in Fig.~\ref{fig:dirac-occ-spliting}(b). Notably, the projected occupation under the transformation K \(\leftrightarrow\) K$'$ is symmetric because inversion symmetry relates these two points, even though time-reversal symmetry has been broken. The quasienergy and the steady-state occupation, however, remain symmetric, as shown in Fig.~\ref{fig:dirac-occ-spliting}(a) and Fig.~\ref{fig:dirac-occ-spliting}(c), because these quantities do not depend on the static eigenstates.
\end{document}